**Effect of annealing temperatures on the electrical conductivity and dielectric properties of $Ni_{1.5}Fe_{1.5}O_4$ spinel ferrite prepared by chemical reaction at different pH values**

K.S. Aneesh Kumar and R.N. Bhowmik[*]

Department of Physics, Pondicherry University, R. Venkataraman Nagar,

Kalapet, Pondicherry-605014, India

[*]Corresponding author: Tel.: +91-9944064547; Fax: +91-413-2655734

E-mail: rnbhowmik.phy@pondiuni.edu.in

**Abstract**

The electrical conductivity and dielectric properties of $Ni_{1.5}Fe_{1.5}O_4$ ferrite has been controlled by varying the annealing temperature of the chemical routed samples. The frequency activated conductivity obeyed Jonscher's power law and universal scaling suggested semiconductor nature. An unusual metal like state has been revealed in the measurement temperature scale in between two semiconductor states with different activation energy. The metal like state has been affected by thermal annealing of the material. The analysis of electrical impedance and modulus spectra has confirmed non-Debye dielectric relaxation with contributions from grains and grain boundaries. The dielectric relaxation process is thermally activated in terms of measurement temperature and annealing temperature of the samples. The hole hopping process, due to presence of $Ni^{3+}$ ions in the present Ni rich ferrite, played a significant role in determining the thermal activated conduction mechanism. This work has successfully applied the technique of a combined variation of annealing temperature and pH value during chemical reaction for tuning electrical parameters in a wide range; for example dc limit of conductivity ~ $10^{-4}$-$10^{-12}$ S/cm, and unusually high activation energy ~ 0.17-1.36 eV.

# 1. Introduction

Spinel ferrites are promising magneto–electronic materials that can be used in the field of electromagnetic devices, power electronics, sprintronics, biomedical applications, MRI scan and hyperthermia treatment [1-4]. The basic needs for applying ferrites in multi-functional devices are soft ferromagnetic properties, flexible electrical conductivity, low dielectric loss, and good chemical stability [5-10]. $NiFe_2O_4$ ferrite is one such material that exhibited tunable electrical properties [11-13] and controlled by the variation of grain size and heat treatment [5, 6, 14-15]. The conduction mechanism in nickel ferrite is controlled by hopping of charge carriers via super-exchange paths $Fe^{3+}$-$O^{2-}$-$Fe^{2+}$ (electron hopping) and $Ni^{2+}$-$O^{2-}$-$Ni^{3+}$ (hole hopping) and significantly affected by the grain and grain boundary structure [5, 16-18]. It is understood that hole hopping ($Ni^{2+}$-$O^{2-}$-$Ni^{3+}$) mechanism dominates at high temperature regime and electron hopping ($Fe^{3+}$-$O^{2-}$-$Fe^{2+}$) mechanism dominates at low temperatures [19]. The increase of Ni content (x >1) in $Ni_xFe_{3-x}O_4$ ferrite increases a considerable amount of $Ni^{3+}$ ions in B sites of the spinel structure. This provides a scope of studying electrical properties in nickel ferrite where hole hopping plays a major role.

Most of the reports [20-22] studied the electric properties of Fe-rich regime (0 ≤ x ≤ 1) of $Ni_xFe_{3-x}O_4$. There is hardly few report on electrical conductivity and dielectric properties of Ni rich regime (1 ≤ x ≤ 2). Hence, we have selected the composition $Ni_{1.5}Fe_{1.5}O_4$, a slightly Ni rich ferrite in comparison to $NiFe_2O_4$, to study the effect of the variation of pH value during chemical reaction and post annealing temperature on different physical properties. The dielectric [23] and room temperature magnetic [24] properties for the $Ni_{1.5}Fe_{1.5}O_4$ samples prepared at different pH value and annealed only at 1000 $^0C$ have been published. The variation of annealing temperature of the as-prepared (chemically routed) sample showed a significant change in micro-structure [25] and it can also lead to a remarkable change in material properties.

In this work, we show the effects of the variation of annealing temperatures on electrical conductivity and dielectric properties of $Ni_{1.5}Fe_{1.5}O_4$ ferrite samples, which were prepared using chemical reaction at pH values in the range 6-12. Secondly, we have analyzed the experimental data using existing models to present a generalized mechanism of the electrical conduction and dielectric properties, and the role of grains and grain boundaries as the function of annealing temperatures of the samples. Finally, we anticipate that this work can be used as a reference to tune electrical properties in different materials by varying the parameters in chemical route.

## 2. Experimental

The material was prepared by chemical reaction of the required amounts of $Ni(NO_3)_2 \cdot 6H_2O$ and $Fe(NO_3)_3 \cdot 9H_2O$ solutions at selected pH values 6, 8, 10, and 12. The chemical reaction of the mixed solution was performed at 80 °C by maintaining nearly constant pH value. The as prepared samples, after washing, were annealed in the temperature range 500 °C -1000 °C with heating rate 5 °C/min. The prepared samples are labeled as NFpHX_Y, where X is the pH value during chemical reaction and Y is the annealing temperature in degree centigrade. The readers can refer earlier work [25] to get details of the material prepared at pH values 8, 10 and 12, and structural phase stabilization by annealing the samples in air. The material prepared at pH value 6 was annealed under high vacuum (~$10^{-5}$ mbar) to get the single phased cubic spinel structure, because impurity hematite phase (α-$Fe_2O_3$) appeared after annealing of the as-prepared sample in air. The X-ray diffraction pattern of all the samples, used in this work, matched to single phased cubic spinel structure with space group Fd3m. The structural phase evolution with variation of annealing temperature has been illustrated for the samples prepared at pH 6 and pH 8 (supplementary Fig.1s). The details of the samples are provided in Table 1. The lattice constant of the samples in cubic spinel phase was found in the range 8.2895-8.3490 A°. Grain size of the samples prepared at pH value 6, 8, 10, and 12 are in the range 15-95 nm, 10- 90 nm, 15-80 nm, and 5-30 nm, respectively. The disc shaped ($\varnothing \sim 12$ mm, t $\sim 1$ mm) samples were used for dielectric measurements in the frequency ($\nu$) range 1-$10^7$ Hz at ac field amplitude 1 V and measurement temperature 173 K-573 K with interval 20 K. The disc shaped samples were sandwiched between two gold coated plates and connected to broadband dielectric spectrometer (Novocontrol, Germany) using shielded cables. WINDETA software was used for acquisition data during dielectric measurement. The magnetic coercivity of the samples at room temperature was calculated from magnetic loop, measured using PPMS (quantum Design, USA).

## 3. Results and discussion

### 3.1. *Analysis of AC conductivity spectra*

Fig. 1 (a-k) shows the frequency ($\nu$) dependence of the real part of ac conductivity ($\sigma'$) at selected measurement temperatures for the samples prepared at specific pH values and annealed at selected temperatures. The $\sigma'(\nu)$ curves at low measurement temperatures (typically below 233 K) followed a linear frequency response and showed small dc limit of conductivity ($10^{-12}$ to $10^{-13}$ S/cm). This is the nearly constant loss (NCL) region of conductivity [26], where conductivity is controlled by localized hopping of ions in an asymmetric double well potential [27]. The increase of measurement temperature (233 K to 573 K) activates the hopping of charge carriers between two ions at lattice sites. In this temperature range, $\sigma'(\nu)$ spectra showed two conductivity regimes (marked by dotted lines in Fig. 1(c)). In regime 1, $\sigma'(\nu)$ is nearly frequency independent and denoted as the dc limit of conductivity ($\sigma_{dc}$). The $\sigma'(\nu)$ curves are frequency activated in regime 2. In addition to these two regimes, the $\sigma'(\nu)$ curves slowly decreased at high frequencies (regime 3) for some of the low measurement temperatures. The σ′(ν,T) spectra in the entire measurement temperature scale obeyed Jonscher's power law with two separate contributions, in addition to dc limit of conductivity [20, 28].

$$\sigma'(\nu, T) = \sigma_{dc}(T) + \sigma_{gb}(T)\nu^{n_{gb}} + \sigma_{g}(T)\nu^{n_g} \qquad (1)$$

The first term $\sigma_{dc}(T)$ arises due to thermal activated transition of electrons from valence band to conduction band in semiconductor ferrites. The second term arises due to short range hopping of charge carriers at the grain boundaries (regime 2). The third term denotes the conductivity at high frequencies (regime3) due to localized hopping of charge carriers within grains. In spinel oxides, the slowly moving electrons under the application of ac electric field distort the electrical charge configuration in lattice structure. A strong electron-phonon coupling between a moving electron and vibrating lattice ions at finite temperature forms a bound charge carrier, known as polaron [29]. The spatial dimension of a polaron extends in the lattice structure of the order of lattice constant for small polarons or beyond the lattice constant for large polarons. Table 1 summarizes the range of $n_{gb}$ and $n_g$ values (within error bar ± 0.01) obtained for measurement temperature range 173 K-573 K for the samples prepared at pH 6, 8, 10 and 12. The exponents $n_{gb}$ and $n_g$ measure the degree of interaction of the charge carriers with surrounded ions while performing hopping between Fe ($Fe^{2+} \leftrightarrow Fe^{3+}$) and Ni ($Ni^{2+} \leftrightarrow Ni^{3+}$) lattice sites. The temperature dependence of $n_{gb}$ (contribution from grain boundaries) and $n_g$ (contribution from grains) (plot is not shown) indicated a local minimum in the temperature range 200 K-500 K, whose position shifted depending on annealing temperature of the samples. The local minimum in $n_{gb}(T)$ and $n_g(T)$ curves, as plotted for the samples annealed at 1000 $^0$C [23], suggest that overlapping large polaron tunnelling (OLPT) mechanism dominates in the thermal activated charge conduction process [29], irrespective of the annealing temperatures.

To verify the universal conductivity mechanism in the $\sigma'(\nu, T)$ data [30-31] with respect to variation of measurement temperature, pH value during material synthesis and post annealing temperature, we have adopted the following scaling approach.

$$\frac{\sigma'(\nu,T)}{\sigma_o(T)} = f\left(\frac{\nu}{\nu_p}\right) \qquad (2)$$

Here, $\nu_p$ is the critical frequency where the conductivity crosses from dc to dispersion of ac conductivity, $\sigma_o$ is the intercept of $\sigma'(\nu)$ curve on $\sigma'$ axis at $\nu \to 0$ and its value is close to $\sigma_{dc}$. The $\nu_p$ and $\sigma_o$ were adjusted to obtain the best scaling of $\sigma'(\nu)$ data. Fig. 2(a-h) shows the overlapped curves measured in the temperature range 173-573 K for the sample prepared at pH values 6, 8, 10 and 12, and each sample was annealed at two different temperatures. Fig. 2(i-j) shows the scaling of $\sigma'(\nu)$ curves at measurement temperatures 313 K and 513 K for the sample prepared at pH 8 and annealed at different temperatures in the range 500-1000 $^0$C. Fig.2(k-l) shows the scaled $\sigma'(\nu)$ curves at measurement temperatures 513 K and 313 K for the samples prepared at different pH values and annealed at 800 $^0$C. We observed that the conductivity data are scaled into a single master curve over a wide range of measurement temperatures for all the samples. The conductivity curves merged into a single master curve throughout the frequency range almost at all measurement temperatures for the samples annealed at higher temperatures. The conductivity data for the samples annealed at low temperature (500 $^0$C) deviated from the master curve at lower measurement temperatures (where NCL region dominates in conductivity curves) and at frequencies lower than $\nu_p$ and some cases, at higher frequencies also. It may be mentioned that micro-structural heterogeneity determine the character of electrical properties, arising from grain boundaries, for the samples annealed at lower temperatures [32]. Fig. 3(a-l) shows the temperature dependence of $\sigma_0$, which was used for scaling of ac conductivity of the samples. The $\nu_p(T)$ curves (not plotted) reproduced the features similar to that of $\sigma_0(T)$ curves.

The $\sigma_0(T)$ curves in Fig. 3(f) indicated three conductivity states S1M1 (high temperature semiconductor state), M1M2 (intermediate metal like state) and M2S2 (low temperature semiconductor state) for all the samples. The conductivity transition temperatures are marked as $T_{SM}$ (S1M1 ↔ M1M2) and $T_{MS}$ (M1M2 ↔ M2S2) while measurement temperature decreased from 573 K to 173 K. The metal like state (negative slope in $\sigma_0(T)$ curves) is more prominent for the samples annealed at low temperatures, irrespective of pH value during chemical reaction of the material preparation. The metal like state shifts to higher measurement temperature and the negative slope of $\sigma_0(T)$ curves is reduced for samples with higher annealing temperature. Such metal like state has been understood [23, 26] to be related to change of thermal activated conduction mechanism and reconfiguration of the electronic spins of transition metal ions like $Fe^{2+}/Fe^{3+}$ and $Ni^{2+}/Ni^{3+}$ ions among the triply degenerate $t_{2g}$ states and less stable $e_g$ states during exchange of Fe and Ni ions between A and B sites of spinel structure [33]. The high magnetic state and high resistance state are associated with the inverse spinel structure of $NiFe_2O_4$, whereas local disorder in B sites due to site exchange of $Ni^{2+}$ and $Fe^{3+}$ ions promotes metal like state [34]. The magnetic spin interactions inside the grain and grain boundary of magnetic particles are affected by heat treatment of the material. We have analyzed the complex impedance ($Z^*(\nu)$) and modulus ($M^*(\nu)$) spectra to understand the grain and grain boundary contributions of the samples as the function of annealing temperature.

*3.2. Analysis of impedance spectra*

Fig. 4 (a-l) shows the Cole-Cole plots in complex impedance diagram (-$Z''$ vs. $Z'$) for the samples prepared at pH values 6-12, and annealed at different temperatures. We noted that the resistance ($R_{el}$) contribution from electrode–sample interface is not significant for the present ferrite system, which generally exhibits a straight line parallel to $Z''$ axis in the impedance plane.

Considering heterogeneous electronic structure in the samples, a constant phase element (CPE) has been included in the equivalent circuit consisting of two parallel R-C elements in series to account for the distribution of relaxation process and ($Z^*(\omega)$) of the samples has been analyzed by following equation [23, 35].

$$Z^*(\omega) = \left(\frac{1}{R_g} + A_g(j\omega)^{\alpha_g}\right)^{-1} + \left(\frac{1}{R_{gb}} + A_{gb}(j\omega)^{\alpha_{gb}}\right)^{-1} \quad (3)$$

The impedance of CPE is given by $Z_{CPE}(\omega) = A^{-1}(j\omega)^{-\alpha}$ where A and α are fit parameters, where A = C (capacitor) with α = 1 for an ideal capacitor and $A^{-1}$ = R (resistance) with α = 0 for an ideal resistor. The parameters $R_g$, $R_{gb}$, $A_g$, $A_{gb}$, $\alpha_g$, and $\alpha_{gb}$ were obtained by fitting the impedance plot at each measurement temperature. The subscripts 'g' and 'gb' represent the contributions from grain (high frequency regime) and grain boundary (low frequency regime), respectively. As shown in Table 1, the $\alpha_{gb}$ values varied in the range 0.98-0.61, where as $\alpha_g$ values varied in the range 1.2-0.83. We have calculated the resistivity (ρ) of the samples by incorporating the dimensions in the expression of resistance (R =ρL/A; A and L are surface area and thickness, respectively). Fig. 5(a-h) shows temperature dependence of resistivity from grains ($\rho_g$) and grain boundaries ($\rho_{gb}$). The values of $\rho_{gb}$ and $\rho_g$ are found in a wide range $10^5$ -$10^{12}$ Ω-cm and $10^4$-$10^{11}$ Ω-cm, respectively. Both $\rho_g$(T) and $\rho_{gb}$(T) curves suggested a transformation of semiconductor ↔ metal like conductivity states with negative temperature coefficient of resistance (NTCR) in semiconductor state and positive temperature coefficient of resistance (PTCR) in metal like state. The $\rho_g$(T) and $\rho_{gb}$(T) curves also showed a shift of the metal like state to higher temperature with the increase of annealing temperature of the samples. But, impedance analysis cannot resolve the contribution of $\rho_{gb}$ and $\rho_g$ for the whole measurement temperature scale. The modulus formalism is the best option to extract the contributions from grains and grain boundaries. This is confirmed from a comparative plot (Fig. 6) of the imaginary part of impedance ($Z''(\omega)$) and electric modulus (M″(ω)) at measurement temperatures 213 K (at low temperature semiconductor state) and 413 K (near to high temperature semiconductor state). Fig.6(a-l) suggests that Z″(ω) peak position ($\nu_{gb}$) of the samples lies well below of 1 Hz at low measurement temperature (213 K) and showed only one peak at lower frequency side for all the samples at 413 K. On the other hand, M″(ω) data showed two peaks or signatures of two peaks in the applied frequency range 1-$10^7$ Hz. The peak at lower frequency side ($\nu_{gb}$) is related to long range hopping of ions from one site to another (slow relaxation) at grain boundaries. The peak at high frequency side ($\nu_g$) is attributed to short range hopping of ions (fast relaxation) confined inside the grains. From physics point of view, a peak like behavior in both M″ (ν) and ε″(ν) suggests that long range hopping (conduction process) coexist with localized charge hopping (dielectric relaxation) in the material [36]. We show a detailed analysis of modulus spectra at all measurement temperatures.

3.3. *Analysis of Electrical modulus spectra*

The complex electrical modulus ($M^*$) is related to the complex dielectric constant ($\varepsilon^*$) and complex impedance ($Z^*$) by the following relations [26, 37].

$$M^* = \frac{1}{\varepsilon^*} = M' + iM'' = i\omega C_o Z^*$$

$$M' = \frac{C_o}{C_g}\left[\frac{(\omega R_g C_g)^2}{1+(\omega R_g C_g)^2}\right] + \frac{C_o}{C_{gb}}\left[\frac{(\omega R_{gb} C_{gb})^2}{1+(\omega R_{gb} C_{gb})^2}\right]$$

$$M'' = \frac{C_o}{C_g}\left[\frac{(\omega R_g C_g)}{1+(\omega R_g C_g)^2}\right] + \frac{C_o}{C_{gb}}\left[\frac{(\omega R_{gb} C_{gb})}{1+(\omega R_{gb} C_{gb})^2}\right] \quad (4)$$

Where $\omega = 2\pi\nu$ is the angular frequency

$C_o = \frac{\epsilon_o A}{d}$ is the empty cell capacitance

$'A'$ is the sample area and $'d'$ is the thickness of the sample. Both $M'(\omega)\ and\ M'(\omega)$ approaches to zero as the applied frequency tend to zero (static limit of dielectric relaxation), as clearly seen at 413 K. This suggests the suppression of interfacial polarization effects in modulus formalism.

The dielectric relaxation information can be obtained by fitting M″(ω) spectra with double peaks in the frequency domain using the modified Bergman proposed function [23].

$$M''(\omega) = \frac{M''_{max}}{1-\beta+\left(\frac{\beta}{1+\beta}\right)\left(\beta\left(\frac{\omega_{max}}{\omega}\right)+\left(\frac{\omega}{\omega_{max}}\right)\right)^\beta}, \omega = 2\pi\nu \qquad (5)$$

The fit of M″(ν) spectra using equation (5) are shown in Fig. 7(a-l) at selected measurement temperatures of the samples prepared at specific pH and annealed at different temperatures. The stretched exponential factor, β, represents the strength of interaction in dielectric material [11, 38-39]. As shown in Table 1, the $\beta_{gb}$ (due to grain boundary contribution) and $\beta_g$ (due to grain contribution) are found in the range (0.45-0.96) and (0.2-0.7), respectively. The β values smaller than 1 ($0 \leq \beta \leq 1$) suggest non–Debye type relaxation. A relatively higher value of $\beta_{gb}$ than corresponding value of $\beta_g$ suggests that electronic interactions between charge carriers inside the grains are comparatively stronger than that in the grain boundaries. We observed an interesting change that $\beta_{gb}$ increases while $\beta_g$ decreases with the increase of annealing temperature. It shows a cross over in the conduction mechanism from grain boundary dominated process to grain dominated process with the increase of annealing temperature of the chemical routed samples.

We defined relaxation time (τ) from the M″(ν) peaks at low frequency ($\tau_{gb}$ for grain boundary) and ($\tau_g$ relaxation time for grains) high frequency, respectively. The temperature dependence of $\tau_{gb}$ and $\tau_g$ are given in Fig. 8(a-d) and Fig. 8(e-h). The temperature variation of relaxation time confirmed the existence of three conduction regimes (semiconductor ↔ metal like ↔ semiconductor) in the measurement temperature scale. The relaxation time decreases with the increase of measurement temperature in semiconductor states. In metal like state, the relaxation time increases with temperature due to scattering of delocalized charge carriers. The relaxation time, $\tau_{gb}$ is nearly two orders of magnitude higher than $\tau_g$. This suggests a long ranged charge hopping process inside the grain boundaries, whereas charge hopping process inside the grains is short ranged type. The temperature dependence relaxation time ($\tau$) follows Arrhenius law: $\tau(T) = \tau_o \exp(\frac{E^a}{k_B T})$ [39]. Here, $E^a$ is the activation energy for charge hopping or relaxation process, $k_B$ is the Boltzmann constant, $\tau_o$ is the high temperature values of $\tau$. The activation energy ($E^a$) was calculated from the slope of $\ln\tau$ vs. 1000/T plots (Fig. 9) in the semiconductor states of the samples. Fig. 9(g-h) shows the variation of activation energy ($E^a_{gb}$, $E^a_g$) with annealing temperature ($T_{AN}$) of the samples calculated for high temperature semiconductor state. The essential feature is that activation energy contributed by grain ($E^a_g$) and grain boundary ($E^a_{gb}$) are found in the range 1.12-0.17 eV and 1.36-0.37 eV, respectively. In this Ni rich ferrite, the general trend for the samples prepared at pH 6 and 8 is that activation energy ($E^a_{gb}$, $E^a_g$) initially increased when $T_{AN}$ increased from 500 °C and after attaining a maximum value at a typical annealing temperature, which also depends on pH values, the activation energy decreases at higher annealing temperatures. The activation energy of the sample prepared at pH 10 showed a monotonic increase with $T_{AN}$ up to 1000 °C. The activation energy of the sample prepared at pH 12 monotonically decreased from 0.84 eV to 0.37 eV for grain boundary conduction and 0.62 to 0.3 eV for grains when $T_{AN}$ increases from 800 °C to 1000 °C. The decreased of activation energy for higher annealing temperatures is consistent to that reported for $NiFe_2O_4$ and similar systems with the increase of annealing temperature [5,16, 39]. Among the prepared samples, the sample prepared at pH 8 showed the largest activation energy for both grain ($E^a_g$: 0.52 eV- 1.1 eV) and grain boundary ($E^a_{gb}$: 0.66 eV -1.36 eV) contributions. The samples prepared at pH 6 showed the smallest activation energy from grains ($E^a_{gb}$: 0.19 eV-0.60 eV). The activation energy obtained for low temperature semiconductor state (not for all the samples) is significantly small in comparison to the values in high temperature semiconductor state. $E^a_g$ is also found to be smaller than the $E_{gb}^a$ for low temperature semiconductor state. For example, $E_g^a \sim 0.32$ eV and $E_{gb}^a \sim 0.35$ eV for NFpH10_800 sample; $E_g^a \sim 0.23$ eV and $E_{gb}^a \sim 0.42$ eV for NFpH12_950 sample; $E_g^a \sim 0.30$ eV and $E_{gb}^a \sim 0.35$ eV for NFpH12_1000 sample. The important point is that activation energy from grain contribution for some of the $Ni_{1.5}Fe_{1.5}O_4$ ferrite samples (prepared at higher annealing temperature) only at high temperature semiconductor state and at low temperature semiconductor state (for all annealing temperatures) matched to the range of activation energy (0.22-0.53eV) reported for $NiFe_2O_4$ ferrite [5, 39]. At this point, we would like to mention a basic difference in the charge hopping mechanism in the present material and $NiFe_2O_4$. The electrical conduction in $NiFe_2O_4$ ferrite is controlled mainly by electron hopping process between Fe ions ($Fe^{2+}$ -O- $Fe^{3+}$) and hole hopping between Ni ions ($Ni^{2+}$- O- $Ni^{3+}$) can have minor role. On the other hand, a greater amount of $Ni^{3+}$ ions is expected in the B sites of Ni rich ferrite ($Ni_xFe_{3-x}O_4$; x > 1) [40]. In our earlier work [24], we have estimated the ionic states and site distribution of Ni and Fe ions in the cubic spinel structure of $Ni_{1.5}Fe_{1.5}O_4$ ferrite using Mössbauer spectroscopy, magnetic data and matching of lattice parameters. The essential information is that ionic states could be assigned as $Fe^{2+}$, $Fe^{3+}$, $Ni^{2+}$, and $Ni^{3+}$ in $Ni_{1.5}Fe_{1.5}O_4$ ferrite and distribution of these ions among A and B sites of the cubic spinel structure depends on the pH value at which the material was chemically prepared. The general tendency on increasing the pH value during chemical reaction is that A site occupancy of $Ni^{2+}$ ions decreases by increasing the amount of $Fe^{3+}$ ions. Subsequently, B site occupancy of $Ni^{2+}$ ions increases by decreasing the equal amount of $Fe^{3+}$ ions, whereas B site occupancy of $Ni^{3+}$ ions kept fixed (0.5) to maintain the overall charge valence state of the structure. This indicates the probability of more hole hopping process ($Ni^{2+}$- O- $Ni^{3+}$) at B sites in the present Ni rich samples. It is well established that activation energy for hole hopping process is larger than electron hopping process in ferrite system [40-41].

Based on a wide difference in the activation energy in the measurement temperature scale, we suggest that charge hopping process at low temperature semiconductor state (M2S2) is dominated by electron hopping process, whereas hole hopping process dominates at high temperature semiconductor state (S1M1). The difference in activation energy between the low temperature and high temperature semiconductor states are used for the reconfiguration of low and high spin states of $Ni^{2+}$ and $Ni^{3+}$ ions, and introduction of local disorder at B sites by thermal activated site exchange of $Ni^{2+}$ and $Fe^{3+}$ ions in spinel structure [33-34]. Subsequently, the system exhibits an intermediate metal like state (M1M2) and electrical conductivity is affected.

**4.** *Summary of the electrical parameters with annealing temperature of the samples*

We present a comparative plot (Fig. 10) for selective electrical parameters with annealing temperature of the samples chemically prepared at specific pH values. Fig. 10(a-d) shows the variation of transition temperatures of conductivity states (grain contribution ($T_{SM}^{g}$, $T_{MS}^{g}$) and grain boundary contribution ($T_{SM}^{gb}$, $T_{MS}^{gb}$)). The low transition temperature ($T_{MS}$) and high transition temperature ($T_{SM}$) showed an increasing trend with annealing temperature of the samples to achieve a maximum (hump) in the annealing temperature range 800 $^0$C-950 $^0$C, depending on the pH value at which the material was chemically prepared. Then, both $T_{MS}$ and $T_{SM}$ decreases at higher annealing temperature of the samples. For some of the samples with low annealing temperature (prepared at pH 6 and 8), the conductivity transition temperatures initially decreased with the increase of annealing temperature. Similar feature was noted in the variation of activation energy with annealing temperature. We attribute such behavior to the grain boundary disorder effect in small grain-sized samples (see Table 1). Fig. 10(e-f) shows the conductivity (inverse of resistivity from impedance analysis) variation at measurement temperature 413 K with annealing temperature of the samples. The conductivity ($\sigma_{gb}$ and $\sigma_g$) curves initially increased with annealing temperature to achieve a maximum at annealing temperature, typically above 950 $^0$C for the samples prepared at pH values 8-12, and above 800 $^0$C for the samples prepared at pH value 6. Then, conductivity again decreases at higher annealing temperatures. The samples prepared at pH 6 have exhibited remarkably high conductivity in comparison to the samples prepared at pH 8-12. Another interesting fact is that conductivity of the samples at any specific annealing temperature showed a general decreasing trend with the increase of pH value from 6 to 12, except some differences in this trend for the samples prepared at pH 8 and 10. The variation of conductivity transition temperatures and conductivity from grain and grain boundary contributions suggest that there is some internal transformation in the electrical conduction process. Such internal transformation depends on annealing temperature of the samples. If we look at the comparative plot (Fig. 6), a wide separation in the positions of $Z''(\omega)$ and $M''(\omega)$ peaks at low frequency side was indicated for low measurement temperature (213 K). It represents a typical localized dielectric relaxation. The difference ($\Delta\nu$) between the positions of $Z''(\omega)$ and $M''(\omega)$ peaks narrowed at higher measurement temperature (413 K), which suggests thermal activated charge conduction process in the material [36]. The variation of $\Delta\nu_{gb}$ (from low frequency peak at 413 K) in Fig. 10 (g) showed an initial increase with annealing temperature of the samples up to 800 $^0$C-950 $^0$C, followed by a noticeable decrease at annealing temperature 1000 $^0$C. This result suggests that the localized character of charge hopping process initially increases with the increase of annealing temperatures. At higher annealing temperature (> 950 $^0$C), the thermal activated charge hopping process increases. We propose that localized hopping of ions in an asymmetric double well potential controls the conductivity mechanism in the low measurement temperature scale (NCL regime), typically below 233 K, whereas thermal activated charge (electrons and holes) hopping process dominates at higher measurement temperatures. The charge hopping due to electrons ($Fe^{2+}$-O-$Fe^{3+}$) is a fast relaxation process in comparison to hole hopping process ($Ni^{2+}$-O-$Ni^{3+}$). The hole hopping process is activated mainly at higher measurement temperature [19]. The fast relaxation contribution of the electrons ($Fe^{2+}$-O-$Fe^{3+}$) may increase significantly in the high temperature semiconductor state of the samples for annealing temperature below 800 °C and above 950 °C, and slow relaxation contribution of holes ($Ni^{2+}$-O-$Ni^{3}$) increases for annealing temperature in the range 800 °C-950 °C. This means increase of the annealing temperature, irrespective of the variation of pH value during chemical reaction, brings a transformation in the conduction mechanism. This is associated with some other intrinsic phenomena, e.g., grain boundary refinement and exchange of Ni and Fe ions among A and B sites of the cubic spinel structure as an effect of thermal annealing of the material. First, we look at the grain boundary refinement effect. It is interesting that the hump like feature in the variation of different electrical parameters is well comparable to the feature of magnetic coercivity at room temperature (Fig. 10(h)). Such typical pattern of the variation of magnetic coercivity is attributed to grain size effect, where magnetic domain structure transforms from single domain (SD) to pseudo-single domain (PSD) to multi-domain (MD) with the increase of grain size or annealing temperature of the samples [8, 17, 24]. Although a detailed Mössbauer study was not performed for the samples prepared at a specific pH value and post annealed at different temperatures, but the results from Raman spectra and FTIR spectra [25] indicated a possible site exchange of cations (Ni, Fe) upon increasing the annealing temperature, in addition to the fact that site distribution of cations in our samples depends on pH value during coprecipitation. For example, the suppression of shoulders below $T_{2g}$ (2) and $A_{1g}$ modes in Raman spectra suggested more B sites population of $Fe^{3+}$ ions by migrating equivalent amounts of $Ni^{2+}$ ions to A sites for the samples prepared at pH 8 and 10 with annealing temperature below 800 $^0$C. Raman spectra also suggested that the samples preferred to be in inverse spinel structure (more population of $Fe^{3+}$ in A sites by exchanging equal amount of $Ni^{2+}$ in B sites) for annealing temperatures $\geq$ 800° C. Such site exchange of Ni and Fe ions is expected to establish a correlation between magnetic and electrical properties in the present ferrites. The present work suggests that the grain boundary refinement process along with contribution from $Ni^{2+} \leftrightarrow Ni^{3+}$ paths in B sites increases the slow charge conduction for the annealing temperature in the range 800 °C-950 °C. The increase of annealing temperature above > 900 °C can increase the population of Fe ions in B sites that increases probability of electron hopping ($Fe^{2+}$-O-$Fe^{3+}$) with fast relaxation and low activation energy in the material [41-42].

## 5. Conclusions

The present work suggests that thermal activated charge conduction in the chemical routed $Ni_{1.5}Fe_{1.5}O_4$ ferrite samples, irrespective of the annealing temperatures, is controlled by overlapping large polaron tunnelling mechanism. We find that ac conductivity data obeyed a universal scaling over a large range of frequency and measurement temperatures for the samples annealed at higher temperatures. The conductivity data deviated from a master curve mainly at lower measurement temperatures and at frequencies lower than $\nu_p$ for the samples annealed at low temperatures. The material exhibited a transformation of conductivity states from low temperature semiconductor state to high temperature semiconductor state with an intermediate metal like state for all the samples at different annealing temperatures. The metal like conductivity state shifts to higher measurement temperature and the negative slope of the $\sigma_0(T)$ curves reduced for the samples with higher annealing temperature. A peak behavior in both $M''$ ($\nu$) and $\varepsilon''$ ($\nu$) suggested a long range hopping (conduction process) coexist with localized charge hopping (dielectric relaxation) in the material. The metal like state is understood as the effect of a crossover of the charge hopping dynamics from localized hopping at lower measurement temperature to thermal activated hopping at higher measurement temperatures. A notably large value of activation energy in the high temperature semiconductor state indicates a significant contribution of hole hopping process through $Ni^{2+}$-O-$Ni^{3+}$ paths. The most attractive result of this work is the establishment of a hump like feature in dc conductivity, activation energy, and conductivity transition temperatures with the variation of annealing temperature. Such feature is understood as a transformation in the conduction process from grain boundary dominated mechanism to grain dominated mechanism with the increase of annealing temperature. The fast relaxation contribution of the electrons ($Fe^{2+}$-O-$Fe^{3+}$) increases significantly in the high temperature semiconductor state for the samples annealed below 800 °C and above 950 °C. The slow relaxation contribution of holes ($Ni^{2+}$-O-$Ni^{3}$) increases for the samples with annealing temperature in the range 800 °C-950 °C.

**Acknowledgment**

The authors thank CIF, Pondicherry University for dielectric measurements. RNB thanks to UGC for supporting research Grant (F.No. 42-804/2013 (SR)) for the present work.

**Figure captions**

Fig. 1(a-k) $\sigma'(\nu)$ curves at selected measurement temperatures for the Ni1.5Fe1.5O4 ferrite synthesized at pH values 6,8,10, and 12, and annealed at different temperatures.

Fig. 2 Scaling of $\sigma'(\nu$ data at different measurement temperatures of the samples (a-h), scaled data at 513 K (i) and 313 K (j) for the samples at pH 8 and annealed at different temperatures, scaled data at 513 K (k) and 313 K (l) for the samples prepared at different pH values in the range 6-12 and annealed at 800 $^0$C.

Fig. 3 Temperature variation of the dc conductivity value ($\sigma_0$) used for scaling of the frequency dependence of ac conductivity curves for different samples.

Fig.4 (a-l) Complex impedance plot for selected samples. Inset of (a) shows the equivalent circuit used for fitting of experimental data (symbol)and lines showed the fit data.

Fig. 5 Temperature dependence of grain boundary ($\rho_{gb}$) and grain ($\rho_g$) resistivity contributions from impedance analysis of the samples at different annealing temperatures.

Fig. 6 Frequency dependence of imaginary part of impedance (left-Y axis) and electrical modulus (Right- Y axis) measured at 213 K (a-f) and 413K (g-l).

Fig. 7 Fit of the Imaginary part of modulus spectra using Bergman proposed function at selected measurement temperatures of the samples prepared at specific pH value and annealed at different temperatures. Lines guide to the fit data and symbol represents experimental data.

Fig. 8 Temperature dependence of relaxation time $\tau$ ($\tau_{gb}, \tau_g$) calculated from imaginary part of modulus spectra for $Ni_{1.5}Fe_{1.5}O_4$ samples annealed at different temperatures.

Fig. 9 Fit of the $\tau_{gb}(T)$ and $\tau_g(T)$ data to obtain the activation energy. The activation energy for grain boundary ($E^a_{gb}$) and grain ($E^a_g$) contributions are shown in (g-h).

Fig. 10 Variation of the conductivity transition temperatures ($T_{MS}$, $T_{SM}$) (a-d), conductivity ($\sigma_g$, $\sigma_{gb}$) at 413 K contributed from grains (e) and grain boundaries (f), difference of the positions in low frequency $Z^{//}(\nu)$ and $M^{//}(\nu)$ peaks (g), and magnetic coercivity at 300 K (h) with annealing temperature of the samples prepared at specific pH samples.

Table. 1. Power law fitted exponent ($n_{gb}$, $n_g$), equivalent circuit fitted CPE parameter ($\alpha_{gb}$, $\alpha_g$), and stretched exponential parameter ($\beta_{gb}$, $\beta_g$) for $Ni_{1.5}Fe_{1.5}O_4$ ferrite system.

| pH value | $T_{AN}$ (°C) | Sample code | Grain size (nm) | $n_{gb}(\pm 0.01)$ | $n_g(\pm 0.01)$ | $\alpha_{gb}(\pm 0.01)$ | $\alpha_g(\pm 0.01)$ | $\beta_{gb}(\pm 0.01)$ | $\beta_g(\pm 0.01)$ |
|---|---|---|---|---|---|---|---|---|---|
| 6 | 500 | NFpH6_500 | 16 | 1.07 – 0.84 | 1.63 – 1.12 | 0.81 – 0.91 | 0.87 – 0.98 | 0.52 – 0.94 | 0.37 – 0.83 |
|  | 600 | NFpH6_600 | 25 | 1.05 – 0.73 | 1.55 – 1.08 | 0.81 – 0.97 | 0.54 – 0.98 | 0.52 – 0.96 | 0.46 – 0.78 |
|  | 800 | NFpH6_800 | 67 | 1.61 – 0.73 | 1.55 – 1.08 | 0.9 – 0.98 | 0.68 – 0.83 | 0.57 – 0.96 | 0.31 – 0.67 |
|  | 1000 | NFpH6_1000 | 95 | 1.02 – 0.72 | 2.12 – 1.07 | 0.86 – 0.93 | 0.89 – 1.01 | 0.5 – 0.91 | 0.22 – 0.38 |
| 8 | 500 | NFpH8_500 | 10 | 1.04 – 0.36 | 1.99 – 0.79 | 0.58 – 0.93 | 0.88 – 0.97 | 0.89 – 0.59 | 0.28 – 0.44 |
|  | 600 | NFpH8_600 | 12 | 1.07 – 0.83 | 2.08 – 0.96 | 0.86 – 0.92 | 0.73 – 0.9 | 0.64 – 0.84 | 0.35 – 0.51 |
|  | 800 | NFpH8_800 | 33 | 0.99 – 0.85 | 1.82 – 1.12 | 0.84 – 0.93 | 0.7 – 0.98 | 0.48 – 0.89 | 0.33 – 0.52 |
|  | 950 | NFpH8_950 | 76 | 0.83 – 0.67 | - | 0.83 – 0.93 | 0.7 – 0.98 | 0.44 – 0.68 | 0.38 – 0.55 |
|  | 1000 | NFpH8_1000 | 92 | 0.98 – 0.76 | - | 0.89 – 0.94 | 0.94 – 1.12 | 0.55 – 0.8 | - |
| 10 | 600 | NFpH10_600 | 15 | 0.94 – 0.8 | 1.2 – 0.97 | 0.84 – 0.93 | 0.87 – 1.1 | 0.62 – 0.96 | 0.45 – 0.5 |
|  | 800 | NFpH10_800 | 48 | 0.99 – 0.7 | 1.3 – 0.99 | 0.78 – 0.99 | 0.82 – 1.14 | 0.46 – 0.99 | 0.36 – 0.68 |
|  | 950 | NFpH10_950 | 49 | 0.92 – 0.62 | 1.3 – 1.07 | 0.87 – 0.96 | 0.87 – 0.94 | 0.5 – 0.98 | 0.3 – 0.67 |
|  | 1000 | NFpH10_1000 | 75 | 1.19 – 0.71 | - | 0.94 – 0.97 | 0.93 – 0.97 | 0.88 – 0.97 | 0.44 – 0.77 |
| 12 | 800 | NFpH12_800 | 6 | 0.69 – 0.97 | 0.94 – 1.23 | 0.86 – 0.91 | 0.6 – 0.9 | 0.64 – 0.77 | 0.28 – 0.55 |
|  | 950 | NFpH12_950 | 14 | 0.79 – 0.97 | 1.01 – 1.4 | 0.84 – 0.97 | 0.71 – 0.91 | 0.48 – 0.81 | 0.44 – 0.5 |
|  | 1000 | NFpH12_1000 | 29 | 0.93 – 1.15 | - | 0.93 – 0.96 | 0.73 – 1.02 | 0.87 – 0.96 | 0.43 – 0.48 |

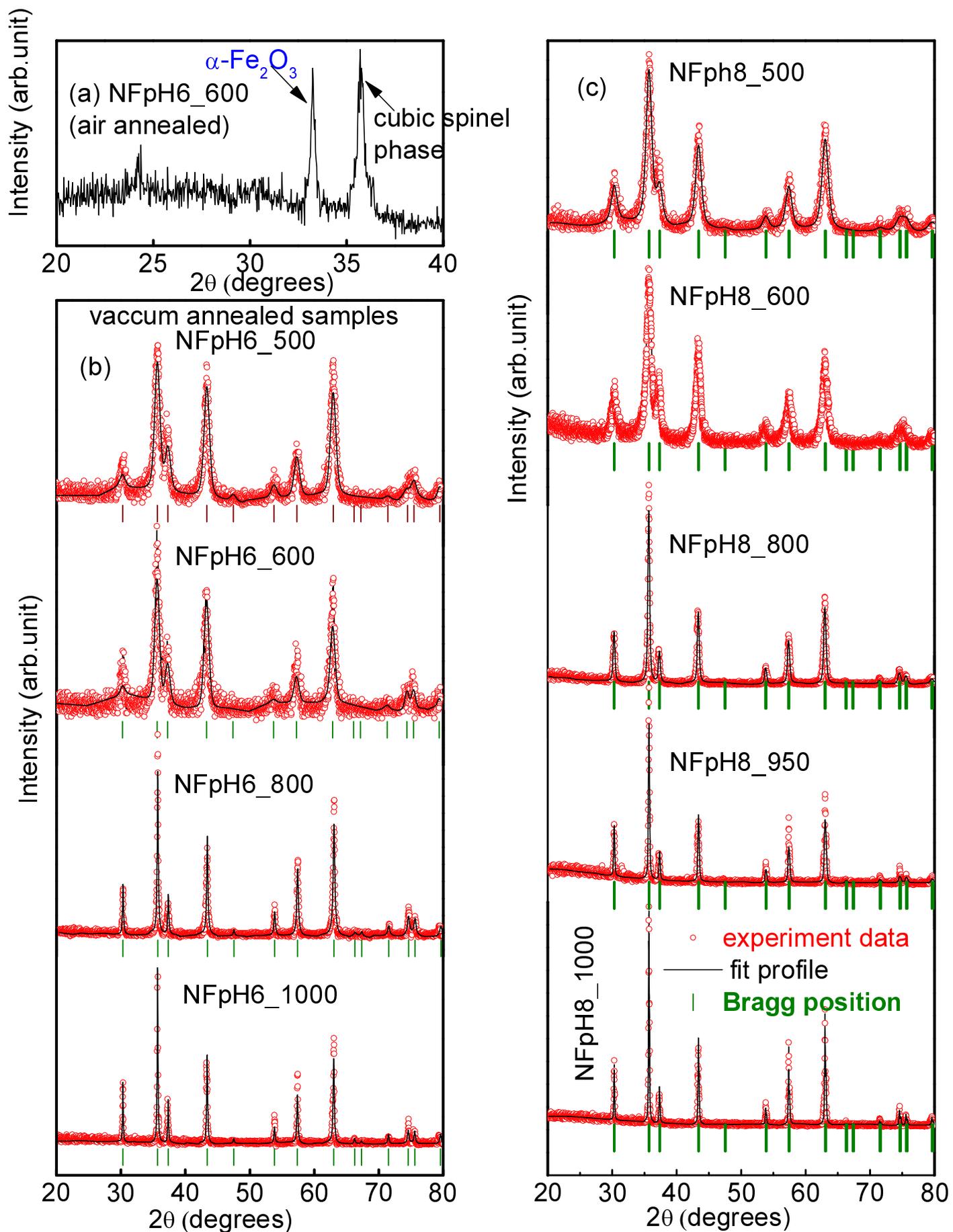

Fig. 1s (a) XRD pattern of the NFpH6_600 (air annealed) with impurity hematite phase. Reitveld profile fit to the XRD pattern for the samples prepared at pH 6 and annealed under vacuum (b) and for the samples prepared at pH 8 and annealed in air (c).

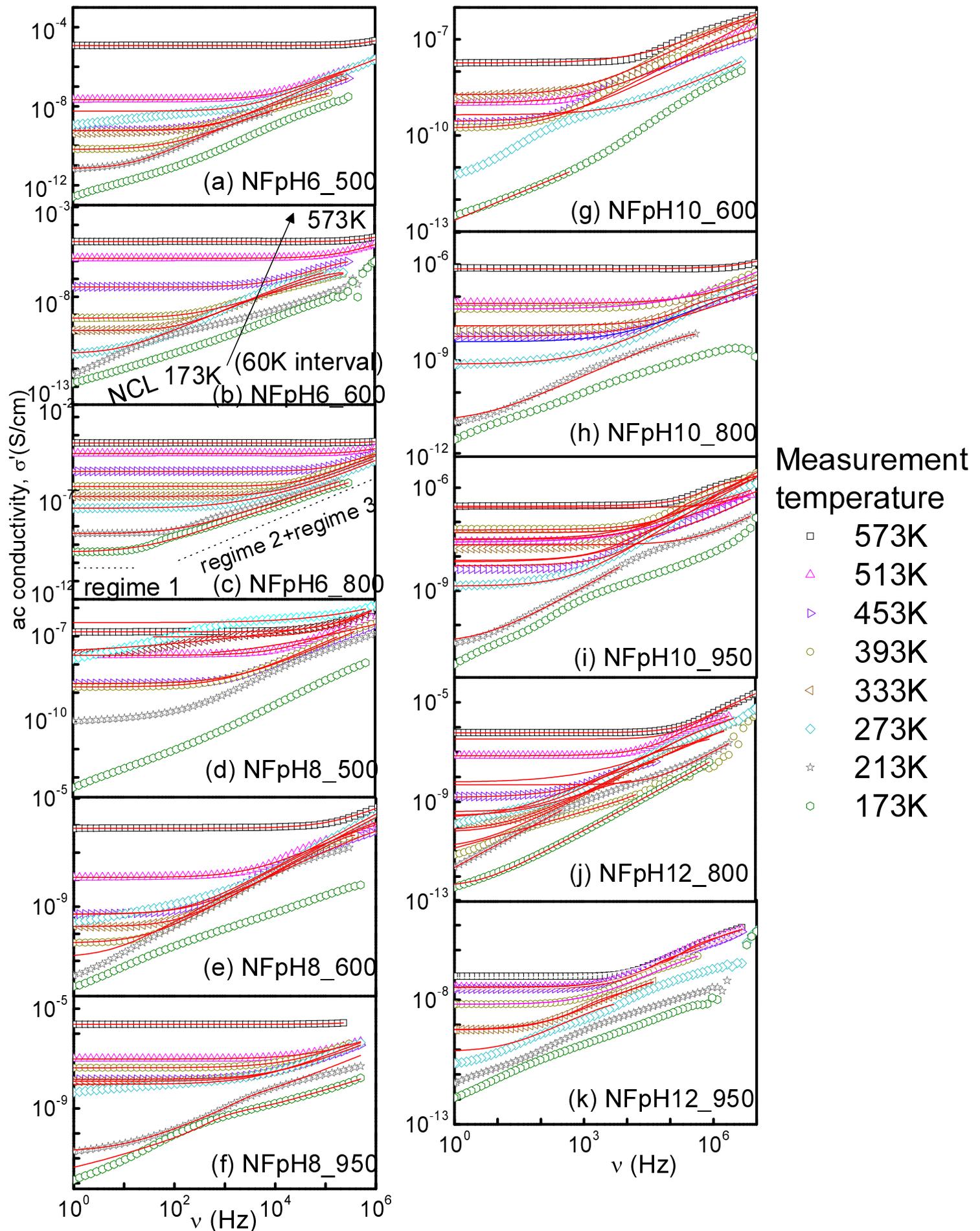

Fig. 1(a-k) $\sigma'(\nu)$ curves at selected measurement temperatures for the $Ni_{1.5}Fe_{1.5}O_4$ ferrite synthesized at pH values 6, 8, 10, and 12, and annealed at different temperatures.

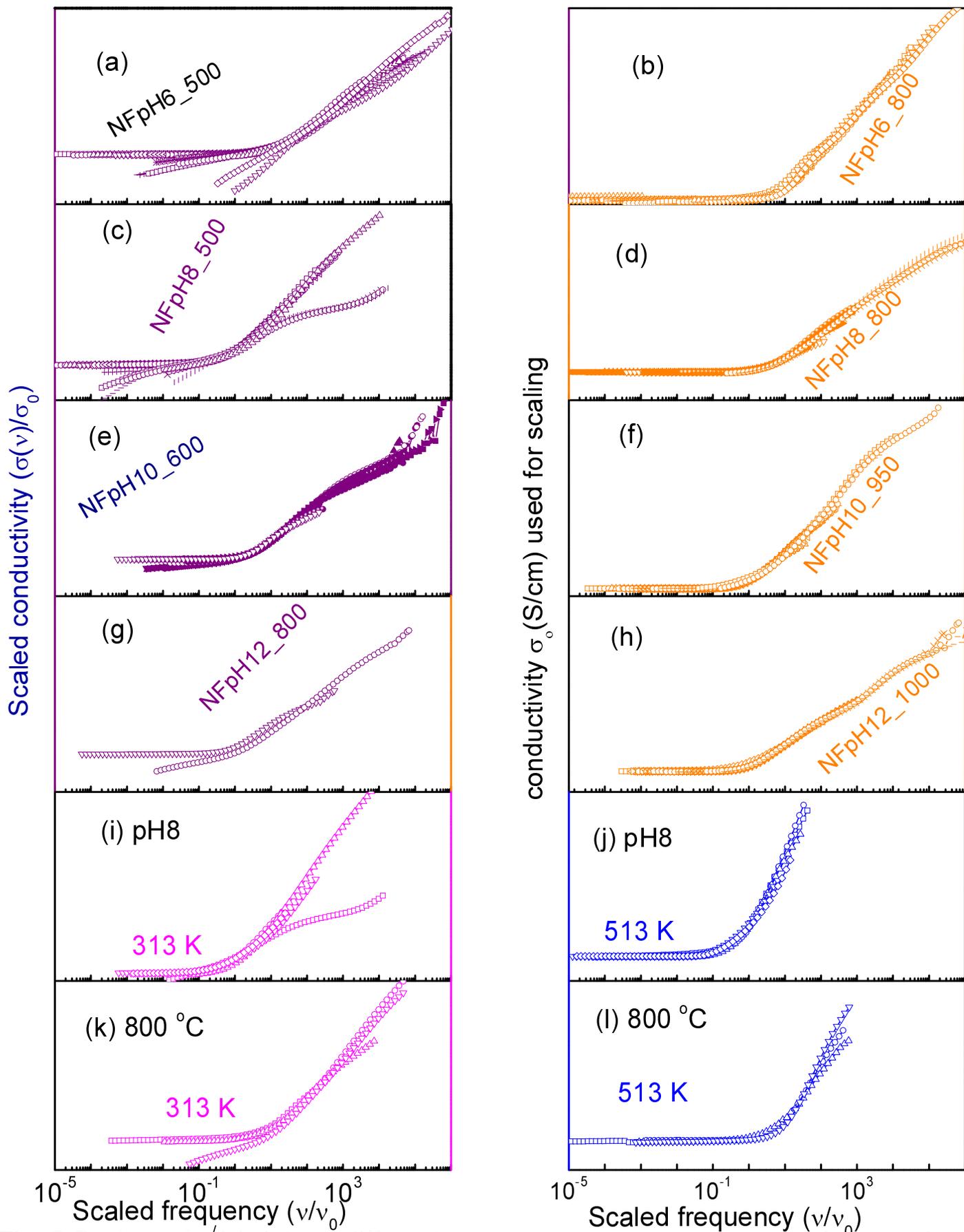

Fig. 2 Scaling of $\sigma'(\nu)$ data at different measurement temperatures of the samples (a-h), scaled data at 513 K (i) and 313 K (j) for the samples at pH 8 and annealed at different temperatures, scaled data at 513 K (k) and 313 K (l) for the samples prepared at different pH values in the range 6-12 and annealed at 800 $^{o}$C.

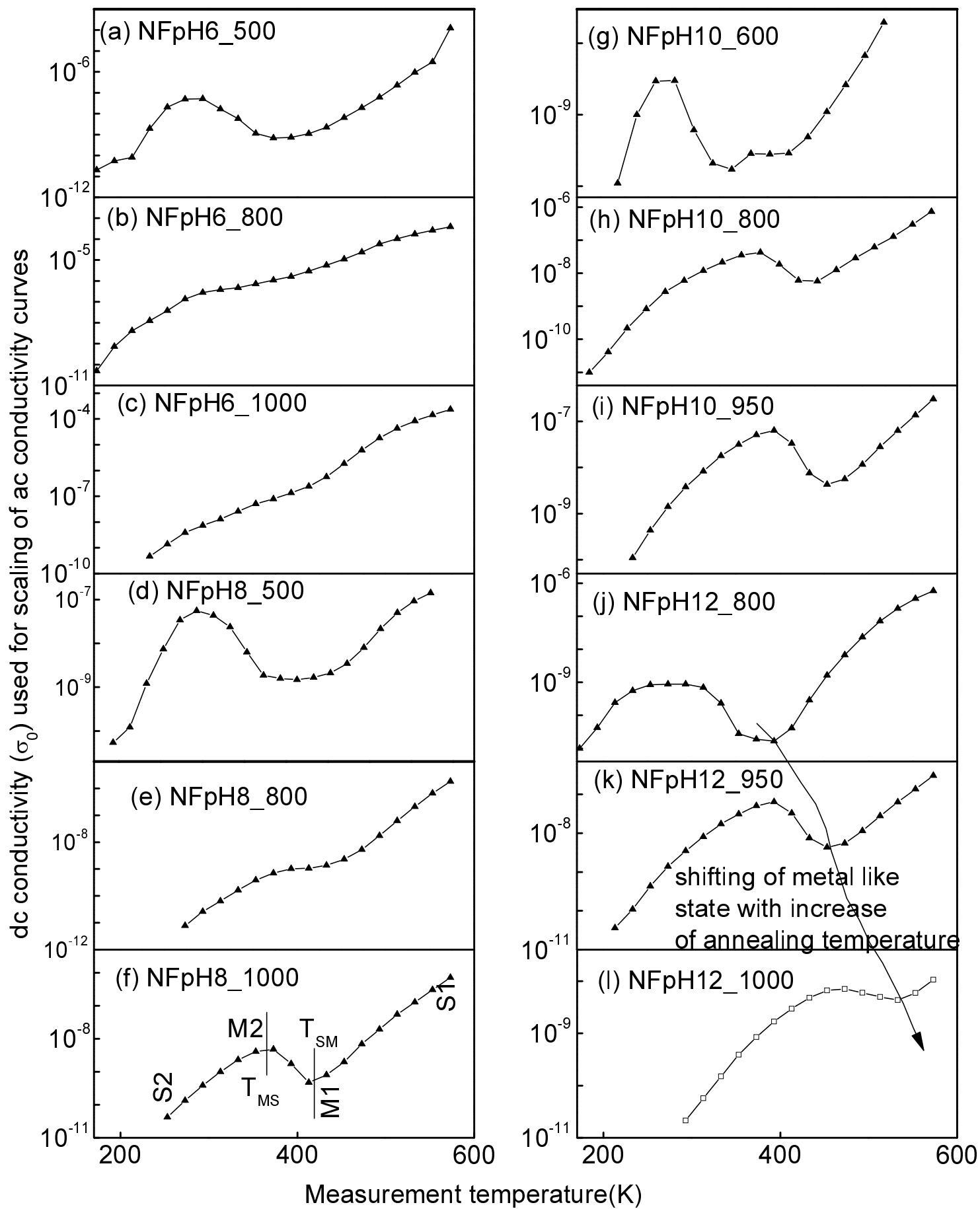

Fig. 3 Temperature variation of the dc conductivity value ($\sigma_0$) used for scaling of the frequency dependence of ac conductivity curves for different samples.

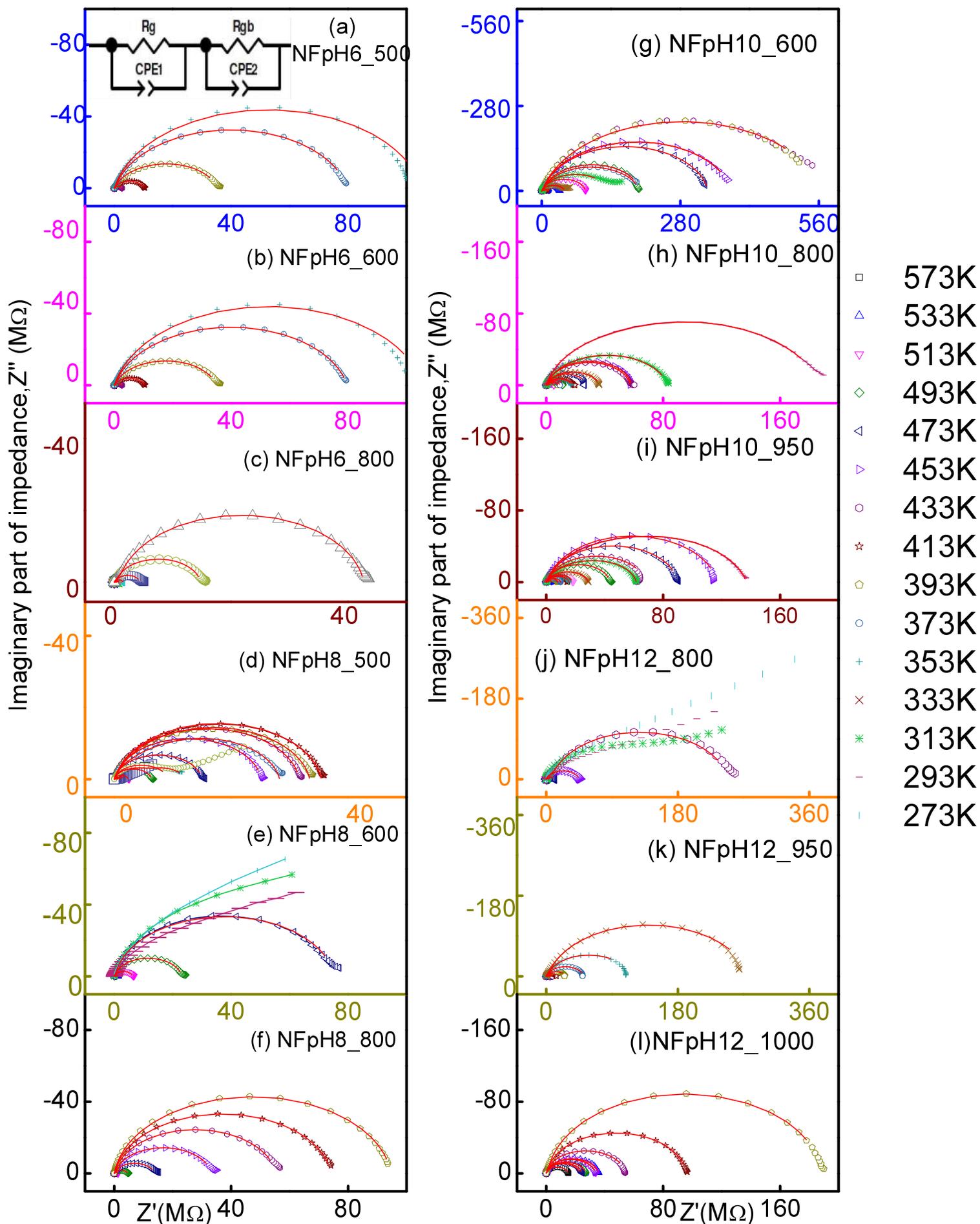

Fig. 4(a-l) Complex impedance plot for selected samples. Inset of (a) shows the equivalent circuit used for fitting of experimental data (symbol) and lines showed the fit data.

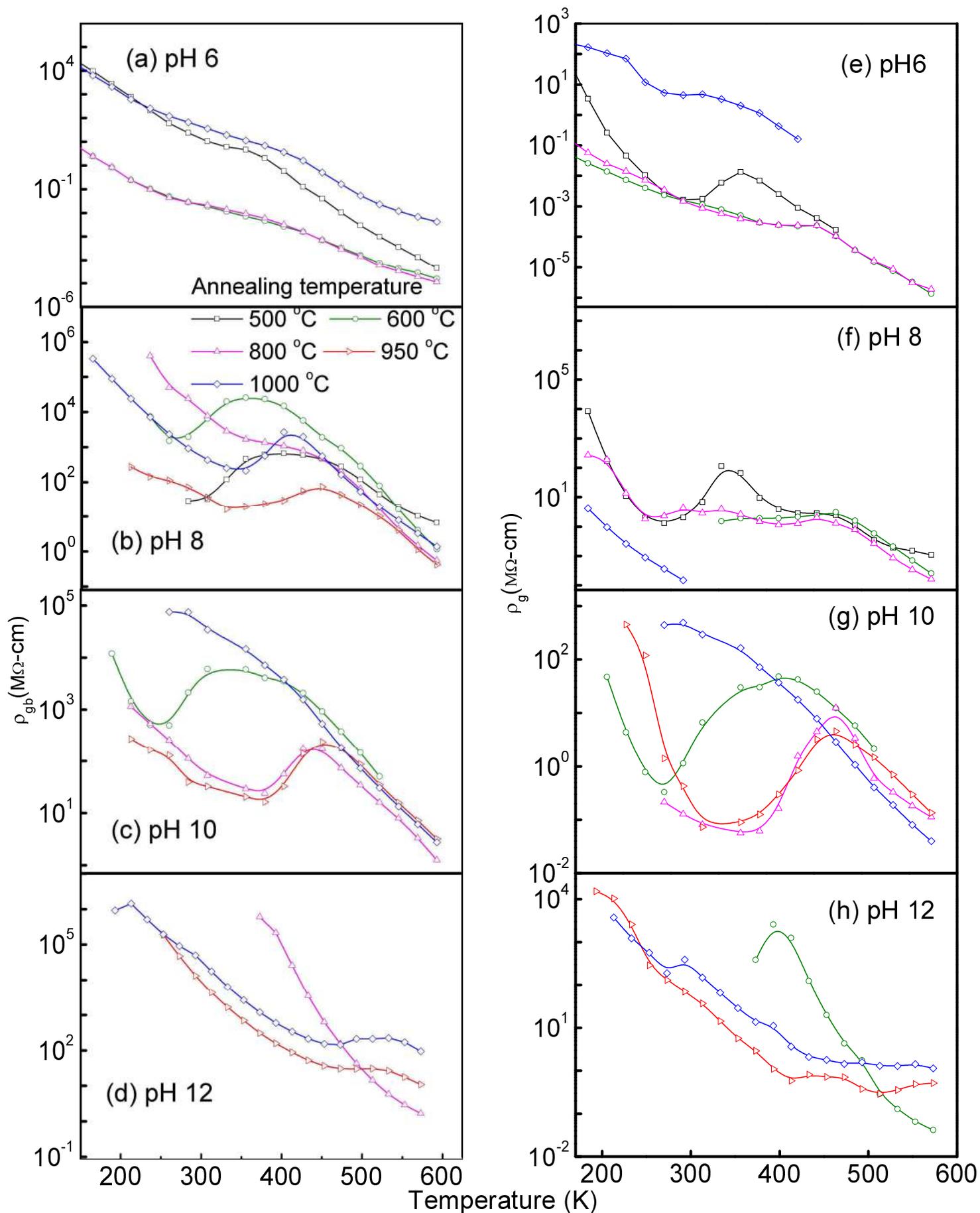

Fig. 5 Temperature dependence of grain boundary ($\rho_{gb}$) and grain ($\rho_g$) resistivity contributions from impedance analysis of the samples at different annealing temperatures

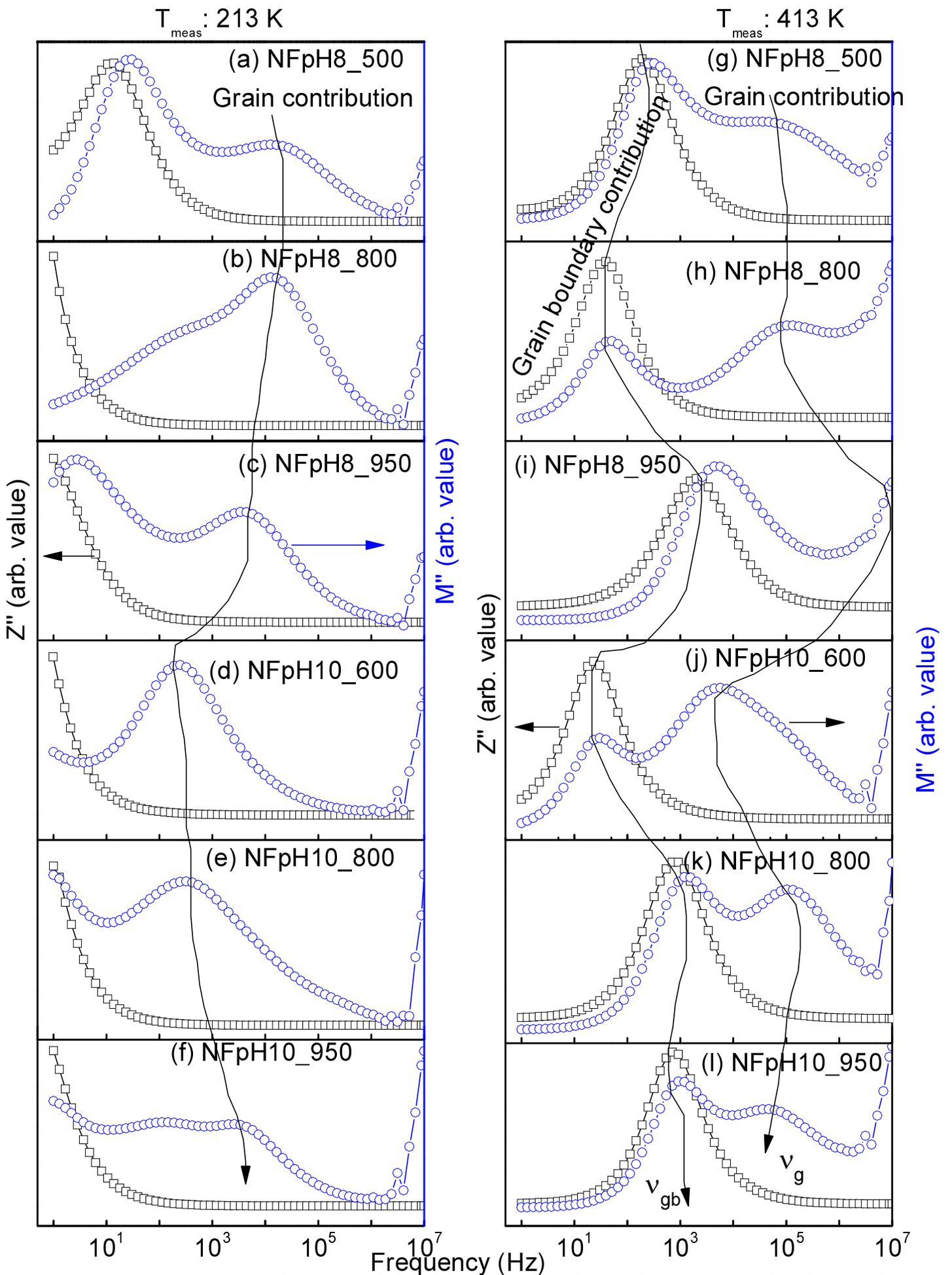

Fig. 6 Frequency dependence of imaginary part of impedance (left-Y axis) and electrical modulus (Right-Y axis) measured at 213 K (a-f) and 413 K (g-l).

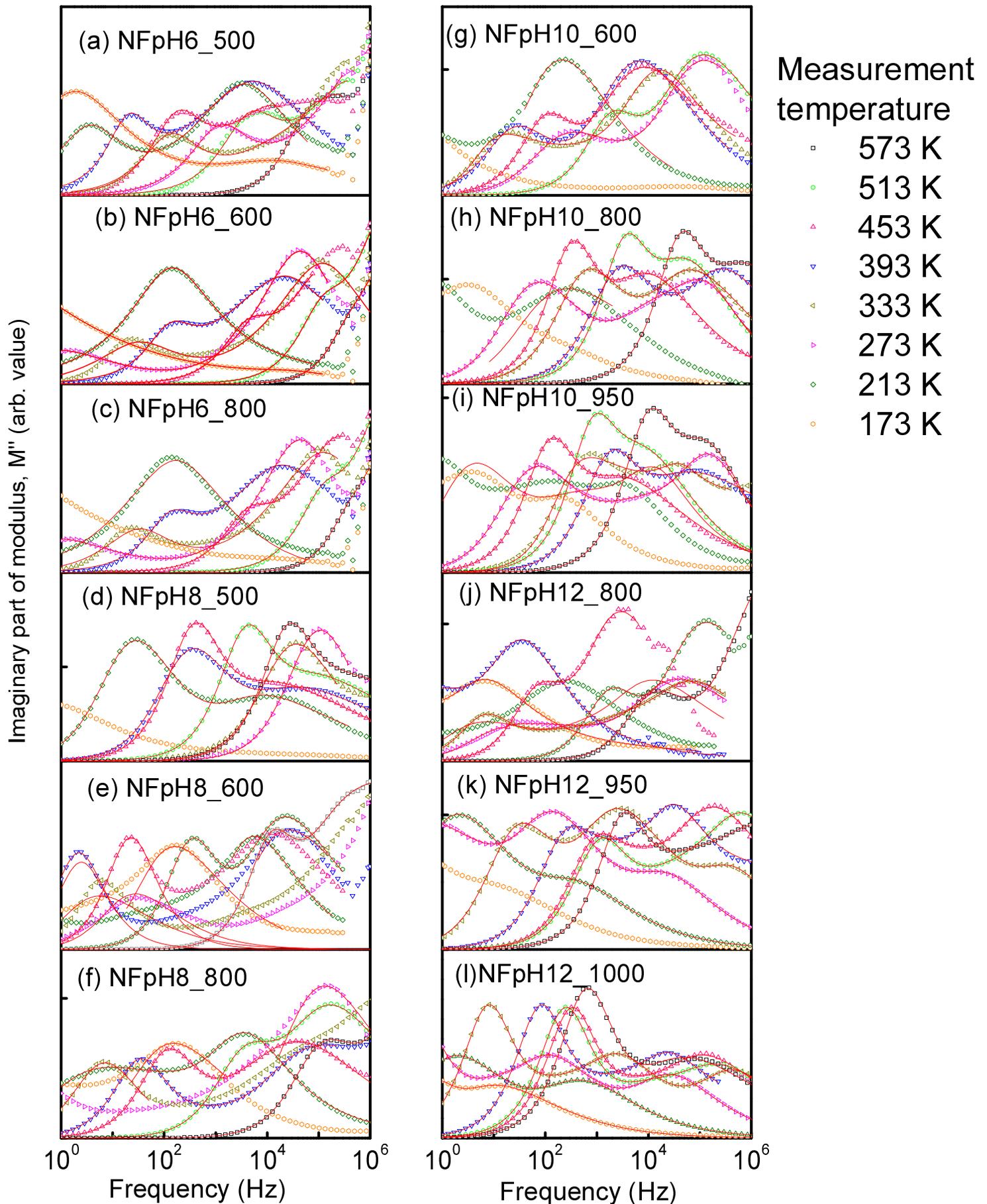

Fig. 7 Fit of the Imaginary part of modulus spectra using Bergman proposed function at selected measurement temperatures of the samples prepared at specific pH value and annealed at different temperatures. Lines guide to the fit of Modulus data and symbol represents experimental data.

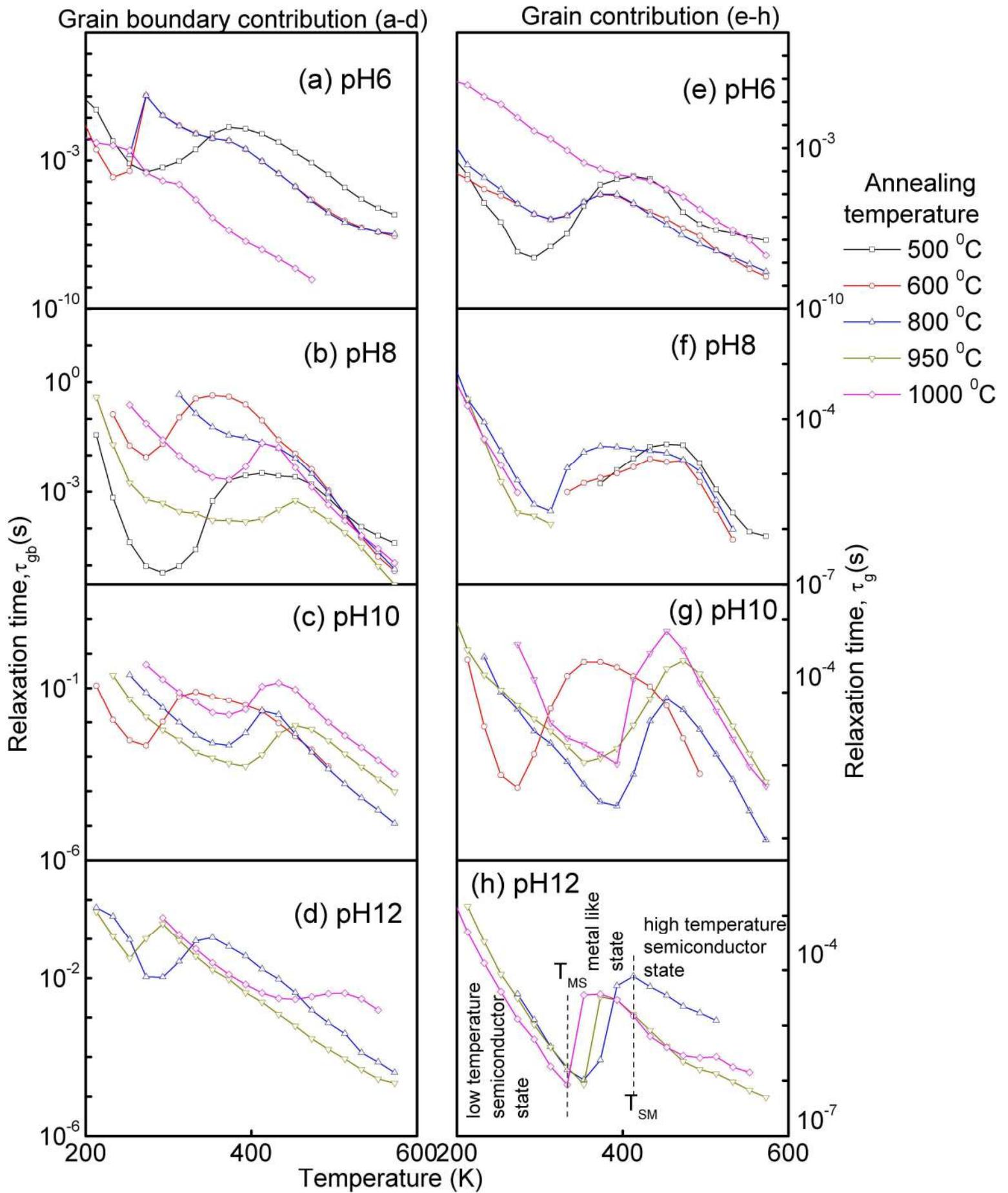

Fig.8 Temperature dependence of relaxation time $\tau$ ($\tau_{gb}$, $\tau_g$) calculated from imaginary part of modulus spectra for $Ni_{1.5}Fe_{1.5}O_4$ samples annealed at different temperatures.

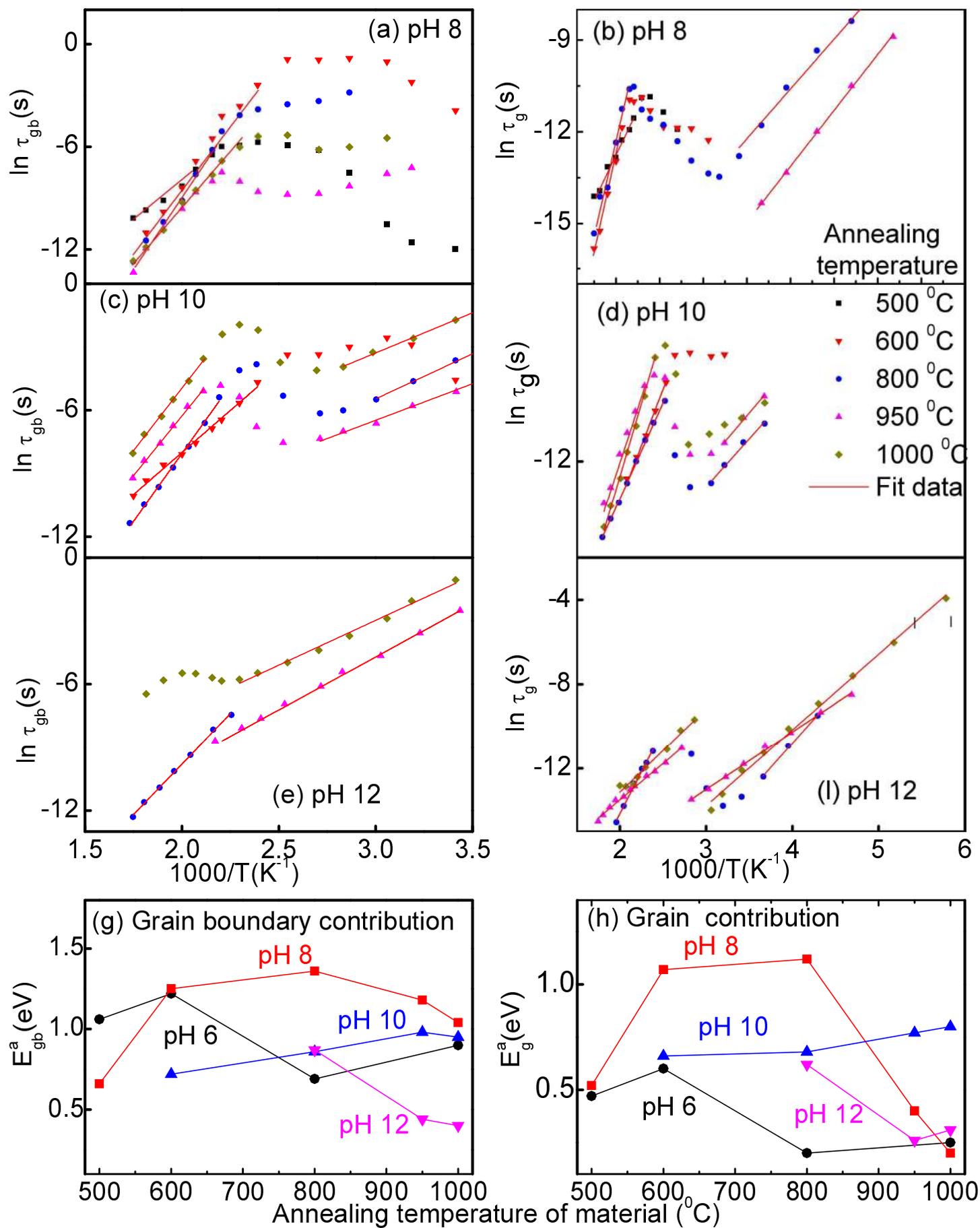

Fig. 9 Fit of the $\tau_{gb}(T)$ and $\tau_g(T)$ data to obtain the activation energy. The activation energy for grain boundary ($E^a_{gb}$) and grain ($E^a_g$) contributions are shown in (g-h).

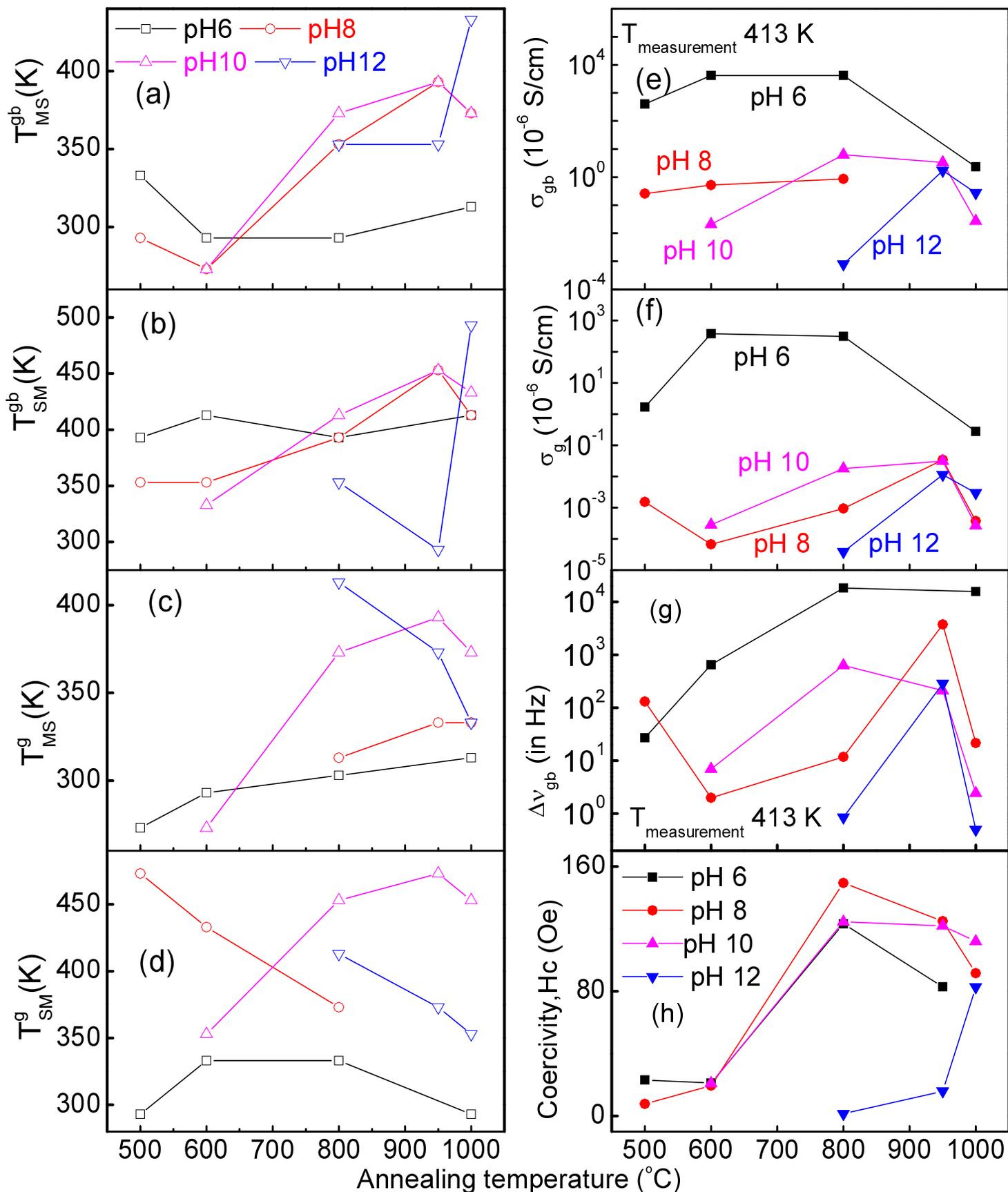

Fig. 10 Variation of the conductivity transition temperatures ($T_{MS}$, $T_{SM}$) (a-d), conducivity ($\sigma_g$, $\sigma_{gb}$) at 413 K contributed from grains (e) and grain boundaries (f), difference of the positions in low frequency $Z''(\nu)$ and $M''(\nu)$ peaks (g), and magnetic coercivity at 300 K (h) with annealing temperature of the samples prepared at specific pH samples.